\documentclass[nofootinbib,amsmath,amssymb,showpacs,showkeys,aps,reprint,onecolumn]{revtex4-1}
\usepackage[utf8,latin1]{inputenc}
\usepackage{graphicx}
\usepackage{dcolumn}
\usepackage[dvipsnames]{xcolor}
\usepackage[T1]{fontenc}

\usepackage{mathrsfs}  
\usepackage{cases}
\usepackage{bm}
\usepackage{academicons}
\usepackage{mathtools, nccmath}
\usepackage{fancyhdr}
\usepackage{tikz,xcolor}
\newcommand{\qm}[1]{``#1''}
\usepackage{tensor}
\usepackage[normalem]{ulem}
\usepackage{lipsum}
\usepackage{soul}
\usepackage{cancel}
\usepackage{stackengine,scalerel}
\usepackage{mathabx}
\usepackage{hyperref}
\usepackage{tabularx}
\hypersetup{colorlinks, linkcolor={red},citecolor={blue},urlcolor={blue}}

\definecolor{lime}{HTML}{A6CE39}
\DeclareRobustCommand{\orcidicon}{
	\begin{tikzpicture}
	\draw[lime, fill=lime] (0,0) 
	circle [radius=0.16] 
	node[white] {{\fontfamily{qag}\selectfont \tiny ID}};
	\draw[white, fill=white] (-0.0625,0.095) 
	circle [radius=0.007];
	\end{tikzpicture}
	\hspace{-2mm}
}

\newcommand{\R}{\mathcal{R}}

\fancypagestyle{plain}{%
  \fancyhf{}
  \fancyfoot[C]{\iffloatpage{}{\thepage}}
  }
\foreach \x in {A, ..., Z}{%
	\expandafter\xdef\csname orcid\x\endcsname{\noexpand\href{https://orcid.org/\csname orcidauthor\x\endcsname}{\noexpand\orcidicon}}
}

\begin{document}

\title[Generalized uncertainty principle corrections in Rastall-Rainbow Casimir wormholes]{Generalized uncertainty principle corrections in Rastall-Rainbow Casimir wormholes}

\author{Emmanuele Battista\orcidB{}$^{1,2,3}$} \email{ebattista@na.infn.it}
\email[\\]{emmanuelebattista@gmail.com}
\author{Salvatore Capozziello\orcidC{}$^{1,2,4}$}
\email{capozziello@na.infn.it}
\author{Abdelghani Errehymy\orcidA{}$^{5}$}
\email{abdelghani.errehymy@gmail.com}

\affiliation{
$^1$ Dipartimento di Fisica ``Ettore Pancini'', Complesso Universitario  di Monte S. Angelo, Universit\`a degli Studi di Napoli ``Federico II'', Via Cintia Edificio 6, 80126 Napoli, Italy,\\
$^2$ Istituto Nazionale di Fisica Nucleare, Sezione di Napoli, Complesso Universitario 
di Monte S. Angelo, Via Cintia Edificio 6, 80126 Napoli, Italy,\\
$^3$Quantum  Theory Center ($\hbar$QTC) \& D-IAS, Southern Denmark University, Campusvej 55, 5230 Odense M, Denmark, \\
$^4$ Scuola Superiore Meridionale, Largo San Marcellino 10, 80138 Napoli, Italy,\\
$^5$ Astrophysics Research Centre, School of Mathematics, Statistics and Computer Science, University of KwaZulu-Natal, Private Bag X54001, Durban 4000, South Africa.
}

\date{\today}

\begin{abstract}

We explore wormhole solutions  sourced by Casimir energy density involving generalized uncertainty principle corrections within the framework of Rastall-Rainbow gravity. The questions of traversability and stability, as well as the presence of exotic matter, are carefully investigated. In particular, the stability issue is addressed via an approach that has not been previously employed  in the context of wormholes. This method, which represents an  improved version of the so-called Herrera cracking technique,  has the potential to yield novel insights in the field of  wormhole geometries.

\end{abstract}

\maketitle

\section{Introduction}

Wormholes (WHs) are fascinating objects predicted by general relativity (GR)  that manifest as tunnel-like structures connecting two distinct spacetime regions.   Spurred by the possibility discussed in the  seminal paper by Morris and Thorne of enabling human interstellar travel \cite{Morris1988cz}, WH solutions are now  widely studied in the literature \cite{visser1995lorentzian,Lobo:2007zb,Alcubierre:2017pqm}. Two key aspects define the WH structures. The first distinctive feature is the {\it throat},  a minimal surface area whose characteristics are crucial for understanding the WH physics and stability. Additionally, the presence  of matter fields that violate the  null energy condition (NEC),  referred to as {\it exotic matter}, which is essential for keeping  WH mouths open and hence ensuring the {\it traversability}. This requirement is linked to the well-known {\it flaring-outward} condition, which imposes precise constraints on the properties of the matter sourcing the WH. 

Various WH models have been devised in the literature attempting to avoid or, at least, reduce the amount of exotic matter required to support the WH, or to find novel and alternative explanations for its origin. For instance,  in  thin-shell WHs \cite{Visser:1989kh,Visser:1989kg,Poisson:1995sv,Lobo:2003xd,DeFalco:2020afv,deCelis:2022ngm,Godani:2023tcx}  exotic matter is concentrated within a surface layer localized at the throat, while phantom WHs \cite{Lobo:2005us,Sushkov:2005kj,Gonzalez:2009cy,Sahoo:2018kct} take into account a phantom energy equation of state (i.e., $ w <-1$, $w$ being the ratio between  the pressure and the energy density) to construct and sustain a traversable WH. This latter framework has been further expanded to include WHs endowed with generalized Chaplygin gas \cite{Lobo:2005vc,Sharif:2014opa,Ghosh:2021dgm} and polytropic phantom energy \cite{Jamil:2010ziq,Cataldo:2013sma,Parsaei:2019utg}. Moreover, recent literature has explored novel categories such as replica WHs \cite{Penington:2019kki,Almheiri:2019qdq,Calmet:2024tgm,Geng:2024xpj}, which are supposed to play a significant role  in the black hole information problem within  the context of Euclidean quantum gravity,  
braneworld WHs \cite{Lobo:2007qi,Wong:2011pt,Sengupta:2021wvi}, which borrow ideas and techniques from brane cosmology, and an intriguing kind of WHs constructed via the so-called  black bounces spacetimes \cite{Simpson:2018tsi,Lobo:2020ffi,Bronnikov:2022bud}, which broaden the class of regular black holes \cite{Lan:2023cvz}.

A  natural scenario  for the existence of WHs is provided by extended or alternative   gravity paradigms  \cite{Capozziello:2011et,Bajardi:2021lwp, CANTATA:2021asi}. These include a plethora of  WH candidates,   ranging from  $f(R)$  \cite{Capozziello:2020zbx,DeFalco:2021ksd,Nath:2023qms,Magalhaes:2023har}, $f(T)$  \cite{Jamil:2012ti},   $f(Q)$ \cite{Banerjee2021a,Hassan:2022hcb,Rastgoo:2024udl}, and $f(R,T)$   \cite{Moraes:2017mir,Rosa:2023guo,Tangphati2023c} gravity models, to hybrid metric-Palatini gravity \cite{Capozziello:2012hr,KordZangeneh:2020ixt,Rosa:2021yym,DeFalco:2021klh}, Einstein-Gauss-Bonnet  \cite{Bhawal:1992sz,Maeda:2008nz,Kanti:2011jz,Kanti:2011yv,Mehdizadeh:2015jra,Panyasiripan:2024iww},  scalar-tensor  \cite{Agnese:1995kd,Bronnikov:2006pt,Franciolini:2018aad},  Einstein-Cartan theories \cite{Bronnikov:2015pha,Mehdizadeh:2017dhb,DiGrezia:2017daq},  non-local gravity \cite{Capozziello:2022zoz} and higher dimensions \cite{Deshpande2022,Deshpande2024}. A recently proposed  framework, that is attracting the  scientific community, is the so-called Rastall-Rainbow (RR) gravity  \cite{Mota:2019zln}, which combines Rastall  \cite{Rastall:1972swe} and Rainbow \cite{Magueijo:2002xx} theories. The key idea of the first scenario consists in  introducing a quantity $\lambda$, dubbed the Rastall parameter, which acts  as a coupling constant amending the ordinary stress-energy tensor conservation law $T_{\: \: \mu;\nu}^{\nu}=0$ via the introduction of a  term proportional to the derivative of the Ricci scalar. The Rainbow model calls for two arbitrary functions,  usually denoted by $\Xi$ and   $\Sigma$, which  depend on the energy of  test particles moving through the spacetime and modify  the standard relativistic energy-momentum dispersion relation $E^2-p^2=m^2$ (with $E$, $p$, and $m$ being the energy, momentum and  rest mass of the particle, respectively). By combining the principles of these two models, one finds that the Einstein field equations assume a generalized form in the RR pattern, where  $\lambda$ embodies the coupling between geometry and matter fields, and the underlying spacetime properties are  influenced by energy owing to the contributions of $\Xi$ and   $\Sigma$. In this context, WH solutions have been discovered, proving that their occurrence  is allowed for specific combination of the free RR  parameters and matter equations of state that mitigate the NEC breach \cite{Tangphati:2023nwz,Errehymy:2024sme,Errehymy:2024lhl}.

Considering  that NEC is naturally violated in quantum field theory, a new   prototype called Casimir WH has been put forward in Ref. \cite{Garattini2019}  (see also Refs. \cite{Maldacena:2018gjk,Zubair2023}). This solution exploits the very well-known fact that the negative  Casimir energy density can represent a potential source of exotic matter realizable in a laboratory, thereby enabling the generation of  traversable WHs. Shortly after this proposal, several extensions were conceived.   Among them, we mention Casimir WHs that involve the corrections to the Casimir effect arising from the generalized uncertainty principle (GUP) \cite{Jusufi:2020rpw,Hassan:2022ibc,Sahoo:2023dus,Muniz2024,Chalavadi:2024jje,Rizwan:2024vop}. The GUP is a pivotal component of quantum gravity models inherently equipped with a minimal length factor $\beta$, which revises the Heisenberg uncertainty principle to account for quantum gravitational effects at small scales \cite{Maggiore:1993rv,Scardigli:2003kr,Casadio:2020rsj}. These contributions can significantly affect both the  traversability and stability of Casimir WHs, thus offering valuable hints into how quantum-gravity regimes can influence WH geometries. 

Driven by these considerations, in this paper we deal with a novel class of static and spherically symmetric Casimir WHs incorporating GUP adjustments within the framework of RR gravity. The role of $\beta$ in  the fundamental questions of  traversability,    stability, and  the presence of exotic matter is  carefully assessed. In particular, the WH stability is evaluated by adopting a  method known as  {\it Herrera cracking technique}, which, to the best of our knowledge,  has never been employed in its most recent and revised form in  WH research settings.

The paper is organised as follows. After  outlining the main features of RR gravity in Sec. \ref{sec2},  in Sec. \ref{Sec.III} we describe the Casimir effect featuring  GUP modifications. GUP-corrected Casimir WHs in RR gravity are then examined in Sec. \ref{Sec:Casimir-WHs}. Discussion and final remarks are reported in Sec. \ref{sec4}.

\section{The Rastall-Rainbow gravity}\label{sec2}

The RR  gravity  originates from the fusion of the Rastall  and  Rainbow theories.  We begin the section by first outlining  the Rastall and Rainbow frameworks separately (see Secs. \ref{Sec:Rastall-theory} and \ref{Sec:Rainbow-theory}). Subsequently, we  delve into the combined RR approach in Sec. \ref{Sec:RR-theory}. 

\subsection{The Rastall gravity}\label{Sec:Rastall-theory}

The Rastall gravity \cite{Rastall:1972swe} is  a phenomenological extension of GR featuring a modified   conservation law of the energy-momentum tensor, where the covariant derivative of  $T_{\mu \nu}$ does not vanish, but, instead, it is related to the Ricci scalar $R$ via an undetermined constant $\Bar{\lambda}$. Explicitly, this  revised relation   takes the form 
\begin{equation}
	T_{\: \: \mu;\nu}^{\nu}=\Bar{\lambda}R_{;\mu},
	\label{conserve_eqn}
\end{equation}
and entails the following reformulation of the field equations:
\begin{equation}
  R^{\nu}_{\: \: \mu}-\frac{1}{2}\delta_{\: \:\mu}^{\nu}R= 8\pi G\left(T_{\: \: \mu}^{\nu}-\Bar{\lambda}\delta_{\: \:\mu}^{\nu}R\right),
  \label{eq8}
\end{equation}
which can also be expressed as
\begin{equation}
    R^{\nu}_{\: \: \mu}-\frac{\lambda}{2}\delta_{\: \:\mu}^{\nu}R= 8\pi G T_{\: \: \mu}^{\nu}. 
  \label{eq9} 
\end{equation}
Here,  we have defined  $\bar{\lambda} := \frac{1 - \lambda}{16 \pi G}$,  $\lambda$ representing the Rastall free parameter that incorporates the nonminimal curvature-matter coupling. The standard Einstein equations are thus recovered  when  $\lambda = 1$ (or $\bar{\lambda} = 0$).

\subsection{The Rainbow gravity}  \label{Sec:Rainbow-theory}

The Rainbow gravity \cite{Magueijo:2002xx} stems from the extension to curved spacetimes of the deformed (or doubly) special relativity  introduced in Ref. \cite{Amelino-Camelia:2000stu}. A crucial aspect of this theory is the modification of the standard relativistic dispersion relation, which becomes
\begin{equation}
E^{2} \Xi^{2}(x) - p^{2}\Sigma^{2}(x) = m^{2}, 
\label{eq1}
\end{equation}
where $x := E/E_{p}$ is a  dimensionless ratio between   the energy $E$ of a probe particle and the Planck energy $E_{p}$, while the correction terms $\Xi(x)$ and $\Sigma(x)$ are known as rainbow functions.  In the infrared  limit, where $x$ approaches zero, they  satisfy the conditions  
\begin{equation}
	\lim_{x\rightarrow 0} \Xi(x)=1, \quad \lim_{x \rightarrow 0} \Sigma(x)=1,
	\label{eq2}
\end{equation}
thus restoring   the standard energy-momentum dispersion relation. 

In the Rainbow theory, the  spacetime metric depends on the energy \cite{Magueijo:2002xx} and hence the  field equations assume the generalized form 
\begin{equation}
 R_{\mu \nu}(x)- \frac{1}{2} g_{\mu \nu}(x) R(x) = 8 \pi G(x) T_{\mu \nu}(x),  
 \label{eq10} 
\end{equation}
where $G(x)$ is an  energy-dependent coupling function usually rescaled as  $G(x)=g^2(x) G$, $G$ being the   Newton constant.

\subsection{The Rastall-Rainbow gravity}  \label{Sec:RR-theory}

The RR gravity  combines the concepts of Rastall and Rainbow  theories \cite{Mota:2019zln}. In this  unified formalism, the modified Einstein  equations can be constructed starting from Eq. (\ref{eq9}) upon  considering an energy-dependent metric $g_{\mu \nu}(x)$ and  coupling constant $G(x)$. Therefore, the RR field equations read as 
\begin{equation}
 R_{\mu \nu}(x)- \frac{\lambda}{2} g_{\mu \nu}(x) R(x) = k(x) T_{\mu \nu}(x),
\label{eqa10}
\end{equation}
where $\lambda$ is supposed to be independent of  the energy and $k(x) := 8 \pi G(x)$.  By adding and subtracting the term $(1/2)g_{\mu\nu}R$, the above equation can be written in a more compact way as
\begin{equation}
R_{\mu\nu}-\frac{1}{2}g_{\mu\nu}R=8\pi G\tau_{\mu \nu}, 
     \label{eq13b}
 \end{equation}
which resembles the ordinary Einstein equations with
 \begin{equation}
     \tau_{\mu\nu}:=T_{\mu\nu}-\frac{(1-\lambda)}{2(1 - 2\lambda)}g_{\mu\nu}T,
      \label{eq13c}
 \end{equation}
 denoting an effective energy-momentum tensor. 

In the next section, we  make a small digression on static and spherically symmetric WH solutions in RR gravity. These will be the starting point of our subsequent analysis. Hereafter, we will assume $G(x)=1=G$ for simplicity;  see also Refs. \cite{Mota:2019zln,Tangphati:2023fey}.

 \subsubsection{Wormhole solutions}

The most general line element for a static and spherically symmetric   geometry in RR gravity  can be expressed  as
\begin{align}\label{metric_RR}
    ds^{2}=-\frac{e^{2\Phi(r)}}{\Xi^{2}(x)} dt^{2}+ \frac{dr^2}{\Sigma^{2}(x) \left(1-\frac{b(r)}{r}\right)}+\frac{r^{2}}{\Sigma^{2}(x)}(d\theta^{2}+\sin{\theta}^{2}d\phi^{2}),
\end{align}
where we have employed standard spherical coordinates ($t$, $r$,  $\theta$,  $\phi$). This metric represents a WH spacetime with $\Phi(r)$ and $b(r)$ denoting the {\it redshift}  and {\it shape} functions, respectively, provided certain well-established conditions are met \cite{Morris1988cz}. First of all, a throat connecting two asymptotically flat regions of the spacetime has to exist. The throat corresponds to a minimal surface area and  satisfies the condition $b(r_0) = r_0$. Moreover, the redshift function $\Phi(r)$ has to remain finite throughout the entire spacetime to prevent the presence of  horizons. Last, the shape function is required to satisfy the so-called flaring-outward condition  near  the throat (see Sec. \ref{Sec:flaring-out}).

In order to search for analytical WH solutions, we consider an anisotropic perfect-fluid energy-momentum tensor
\begin{equation}\label{eq11}
T_{\mu\nu}=(\rho+p_t)u_\mu u_\nu+ p_t g_{\mu\nu}-(p_{t}-p_{r}) \chi_{\mu}\chi_{\nu},
\end{equation}
where $u^\mu$ represents  the fluid four-velocity,  $\chi^{\mu}$  the unit radial vector satisfying $\chi_{\mu} \chi^{\mu} = 1$, $\rho = \rho(r)$ the energy density, and $p_r = p_r(r)$ and $p_t = p_t (r)$  the pressure components along the radial and transverse directions, respectively. 
By evaluating  Eq. (\ref{eq13b}) for the WH geometry (\ref{metric_RR}), one obtains the following expressions for the components of the energy-momentum tensor \cite{Tangphati:2023nwz}:
\begin{subequations}
\label{WH-RR-eqs}
\begin{align}
8\pi  \Bar{\rho}&= \frac{b^{\prime}}{r^{2}},  \label{eq14} \\
8\pi  \Bar{p}_{r}&= 2\left(1-\frac{b}{r}\right)\frac{\Phi^{\prime}}{r}-\frac{b}{r^{3}},  \label{eq15} \\
8\pi  \Bar{p}_{t}&= \left(1-\frac{b}{r}\right)\left[\Phi^{\prime\prime}+\Phi^{\prime 2}-\frac{b^{\prime}r-b}{2r(r-b)}\Phi^{\prime}-\frac{b^{\prime}r-b}{2r^2(r-b)} +\frac{\Phi^{\prime}}{r}\right],  \label{eq16}
\end{align}
\end{subequations}
where the prime  denotes the derivative with respect to the radial coordinate $r$. Here,  we have defined the effective energy density  $\Bar{\rho}$ and the effective radial and tangential pressure components $\Bar{p}_{r}$ and $\Bar{p}_{t}$, respectively, as 
\begin{subequations}
\label{bar-rho-p}
\begin{align}
\Bar{\rho} & := \frac{1}{\Sigma^{2}(x)}\left[\gamma_{1}\rho+\gamma_{2}p_{r}+2\gamma_{2}p_{t}\right],\label{eq17}\\
\Bar{p}_{r} & := \frac{1}{\Sigma^{2}(x)}\left[\gamma_{2}\rho+\gamma_{1}p_{r}-2\gamma_{2}p_{t}\right], \label{eq18}\\
\Bar{p}_{t} & := \frac{1}{\Sigma^{2}(x)}\left[\gamma_{2}\rho-\gamma_{2}p_{r}+\gamma_{3}p_{t}\right], \label{eq18b}
\end{align}
\end{subequations}
with
\begin{equation}
\gamma_{1}:=\frac{1-3\lambda}{2(1-2\lambda)}, \qquad \gamma_{2}:=\frac{1-\lambda}{2(1-2\lambda)}, \qquad
\gamma_{3}:=-\frac{\lambda}{1-2\lambda}.
\label{alpha-def}
\end{equation}

It is worth noticing that Eq. \eqref{WH-RR-eqs} agrees with the conventional Tolman-Oppenheimer-Volkoff  equations for a static spherically symmetric anisotropic fluid in the limit $\lambda=\Sigma(x)=1$.

The set \eqref{WH-RR-eqs} gives rise to a coupled system of three differential equations involving five unknowns: $\Phi(r)$, $b(r)$, $\rho(r)$, $p_r(r)$, and $p_t(r)$. Mathematically, this system is indeterminate. To overcome this issue, one typically assumes a specific form for these quantities, thus finding that WHs are typically sourced by exotic matter \cite{Morris1988cz}. In this paper, we will explore  traversable and stable WHs sourced by the negative Casimir energy density encompassing  GUP corrections.  For this reason, in the next section, we will present the Casimir effect including  modifications provided by the GUP.

\section{Casimir effect with generalized uncertainty principle corrections}\label{Sec.III}

The Casimir effect involves two parallel uncharged conductors experiencing an attractive force due to the distortion of the vacuum of the electromagnetic field, which  can be related to the zero-point energy of quantum electrodynamics \cite{Lamoreaux:2005gf}. Its primary relevance lies in  the occurrence of a  negative  energy density $\rho_{CE}$ that can potentially be generated under laboratory conditions \cite{Garattini:2021kca}.  The explicit expression of $\rho_{CE}$ reads as (henceforth, we set $\hbar=1$) 
\begin{equation}
\rho_{CE}(a)=-\frac{\pi^2}{720}\frac{1}{a^4},\label{eq:rho}
\end{equation}
where $a$ denotes the distance between the conductive plates. 

Since the  prediction  originally made by  Casimir, this phenomenon has  been experimentally validated by many researchers \cite{sparnaay1957attractive, Mohideen:1998iz, Bressi:2002fr,vezzoli2019optical, Good:2019tnf}, 
and currently represents a fundamental research topic which is  widely studied in the literature, ranging from cosmology \cite{Leonhardt:2020fdi,Brevik:2010okp} to supergravity and superstring theory \cite{Goncharov:1987tz,Binetruy:1988dw,Grats:2023xti}. Recently, the  Casimir effect has been investigated in quantum gravity patterns featuring a  modification of the standard position-momentum Heisenberg uncertainty principle \cite{Frassino:2011aa}, known as GUP (see e.g. Refs. \cite{Amati:1988tn, Konishi:1989wk, Maggiore:1993rv,Scardigli:1999jh,Capozziello:1999wx,Scardigli:2003kr,Pedram:2012my,Nozari2012a,Chung:2019efe,Casadio:2020rsj,Bhandari:2024xcr} for further details). Explicitly, the  GUP can be written as 
  \begin{equation} \label{GUP-relation-1}
\Delta X \Delta P \geq \frac{1}{2}\left[1+\beta(\Delta P)^2\right],
\end{equation}
where   we have defined, for any operator $\hat{O}$, the variance  
\begin{align}
    \left(\Delta O\right)^2 := \langle \hat{O}^2 \rangle - \langle \hat{O} \rangle^2,
\end{align}
and the factor $\beta$  is a deforming parameter which puts forth the existence of a minimum length scale in the model. Consequently,   the standard  Heisenberg commutator relation  acquires a correction term  \cite{Frassino:2011aa}:
\begin{align} \label{generalized-commutation}
   \left[ \hat {X}, \hat{P}\right] = i \left(1 + \beta \hat{P}^2\right).
\end{align}
In this framework, a crucial role is fulfilled by the so-called maximally localized quantum states $\vert \psi^{ML} \rangle$, which extend the  usual properties of the position eigenstates $\vert X \rangle$. These states minimize the uncertainty $\Delta X$  and are centered around an average position $\langle \psi^{ML} \vert \hat{X} \vert \psi^{ML} \rangle$ with the best possible resolution \cite{CarvalhoDorsch:2011nyb}.  

When considering $n$ spatial dimensions, the generalized commutation relations \eqref{generalized-commutation}  are modified to the form 
\begin{equation} \label{generalized-commutation-2}
\left[\hat{X}_A, \hat{P}_B \right]= i \left[f(\hat{P}^2)\delta_{AB}+g(\hat{P}^2) \hat{P}_A \hat{P}_B \right],
\end{equation}
where the indices $A$ and $B$ range from $1$ to $n$, and the functions  $f(\hat{P}^2)$ and $g(\hat{P}^2)$  are not completely arbitrary as they are subject to some relations  stemming  from the translational and rotational invariance of Eq. \eqref{generalized-commutation-2}. For the general case with $n>1$, different models arise based on the choice of $f(\hat{P}^2)$ and $g(\hat{P}^2)$, leading to distinct constructions of the maximally localized quantum states.  In the literature, two main proposals have been devised to develop these states: the approach pursued by  Kempf, Mangano, and Mann (KMM) \cite{Kempf:1994su}, and the one followed by Detournay, Gabriel, and Spindel (DGS) \cite{Detournay:2002fq}. Due to the different properties assumed by the  $\vert \psi^{ML} \rangle$ states in  these two patterns,  the standard Casimir energy density \eqref{eq:rho} receives a correction factor. This contribution can be traced back to the  presence of the  minimal length scale brought in by the GUP, which limits the resolution of small distances in the spacetime. Therefore, up to first-order terms in $\beta$, the modified  Casimir energy density can be written as \cite{Frassino:2011aa}
\begin{align}\label{eq27}
\rho(a) = -\frac{\pi^2}{720 }\frac{1}{a^4}\left[1+\xi_{i}\frac{\beta}{a^2}\right], \quad ~~\text{where}~~\left\{ \begin{array}{rcl}
\xi_{KMM} &= \pi^2\left[\frac{28+3\sqrt{10}}{14}\right]  
 \\  \\
\xi_{DGS} &= 4\pi^2\left[\frac{3+\pi^2}{21}\right]    
\end{array}\right. ,
\end{align}
depending on whether one employs  the  KMM or the DGS pattern. 

The GUP-corrected Casimir energy density  \eqref{eq27} represents a unique source of exotic matter that can sustain a WH. The resulting  geometry, evaluated within the framework of RR gravity, will be the focus of the next section.

\section{The wormhole solution}\label{Sec:Casimir-WHs}

Although traversable WHs  are theoretically possible in classical GR,  their existence  relies on the presence of exotic matter and the associated possibility of  contravening the NEC \cite{Morris1988cz}. On the other hand, quantum mechanics is the natural arena allowing for the NEC violation, and the Casimir effect, which   inherently enables a negative energy density, represents an example. Therefore, in this section,  we will deal with GUP-corrected Casimir WHs within the framework of RR gravity.  For simplicity, we hereafter consider  zero-tidal-force WHs, for which the redshift function $\Phi(r)$ can be considered constant \cite{Morris1988cz}. We begin our investigation by analyzing the embedding diagrams in Sec. \ref{Sec:Emdedding-Diagrams}. After that, we study WH traversability and stability in Sec. \ref{Sec:stability}, and we calculate the amount of exotic matter required to keep the WH throat open in Sec. \ref{Sec:exotic-matter}. We conclude the section by evaluating the light deflection angle triggered by the WH gravitational field (see Sec. \ref{Sec:light-deflection}). 

\subsection{Embedding diagrams}\label{Sec:Emdedding-Diagrams}

In the hypothesis of zero-tidal-force WHs, the differential system \eqref{WH-RR-eqs} takes  the much simpler form
\begin{subequations}
\begin{align}
b' &= 8 \pi r^2 \bar{\rho},   
 \label{simplified-1} \\
b &= -8 \pi r^3 \bar{p}_r,
\label{simplified-2}\\
r b'- b &=-16 \pi r^3 \bar{p}_t.
\end{align}    
\end{subequations}
These equations lead to the useful identity $\bar{\rho} +\bar{p}_r = -2 \bar{p}_t$, which translates into (cf. Eqs. \eqref{bar-rho-p} and \eqref{alpha-def})
\begin{align}
p_t= -\left(\frac{\gamma_1 + 3 \gamma_2}{2 \gamma_3} \rho +\frac{1}{2}p_r \right),   
\label{pressure-relation}
\end{align}
which, in turn, allows to write, from Eq. \eqref{eq17}, 
\begin{align}
\bar{\rho}= \frac{\rho}{\Sigma^{2}(x) \lambda}.   
\end{align}
In the above formula,  $\rho$ is represented by the GUP-corrected Casimir energy density  (\ref{eq27}), with  the plate separation distance $a$ replaced by the radial coordinate $r$. In this way,  Eq. \eqref{simplified-1} can be readily integrated and yields
\begin{align}
 b(r) = \frac{\pi ^3 \left(\beta  \xi_i +3 r^2\right)}{270 \lambda  r^3 \Sigma^2 (x)} + \mathcal{C},
\end{align}
where the integration constant $\mathcal{C}$ can be determined by imposing the condition $b(r_0) = r_0$, thus obtaining
\begin{align}\label{sh1}
 b(r) =r_0+\frac{\pi ^3 }{270 \lambda  \Sigma^2(x)} \left[3 \left(\frac{1}{r}-\frac{1}{r_0}\right)+\beta  \xi_i  \left(\frac{1}{r^3}-\frac{1}{r_0^3}\right)\right].
\end{align}
Notice that $b(r)/r $ goes to zero at large distances, and hence   the (energy-dependent) asymptotic flatness condition is fulfilled.  
\begin{figure}[hbt!]
    \centering
    \includegraphics[width = 8.75 cm, height=6.75cm]{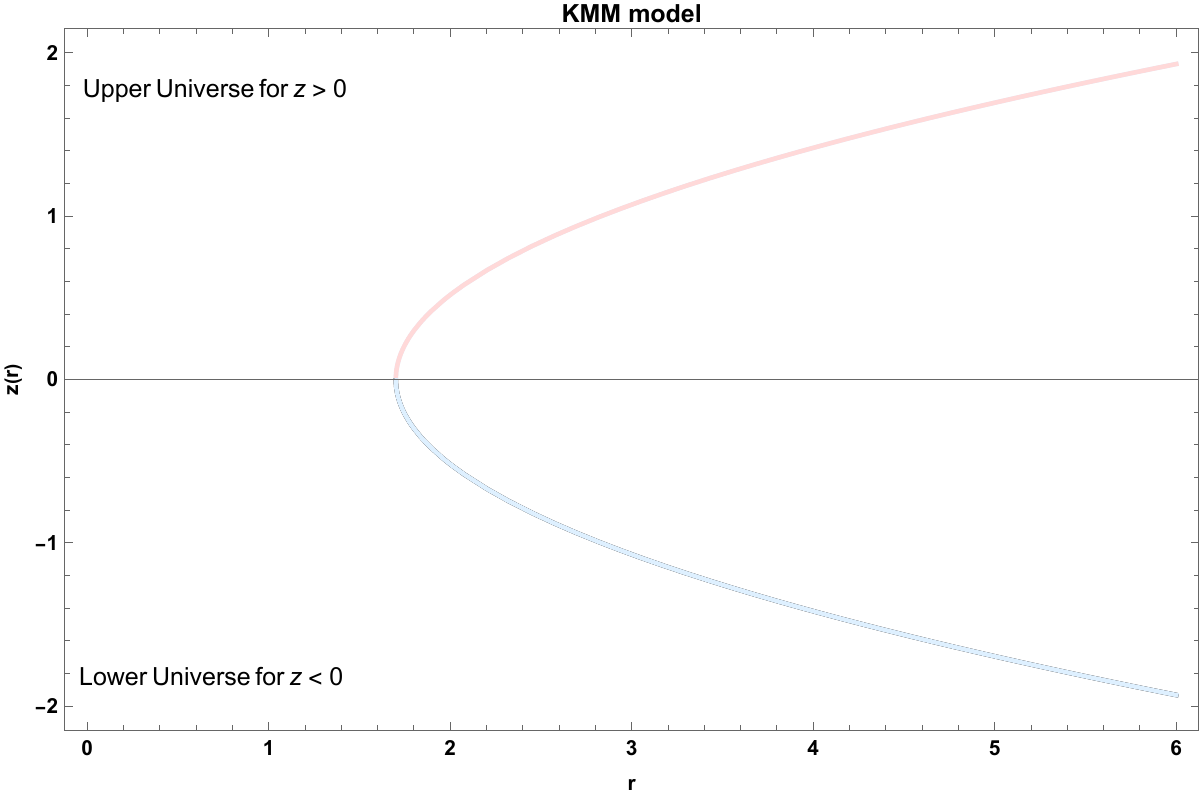}
    \includegraphics[width = 8.75 cm, height=6.75cm]{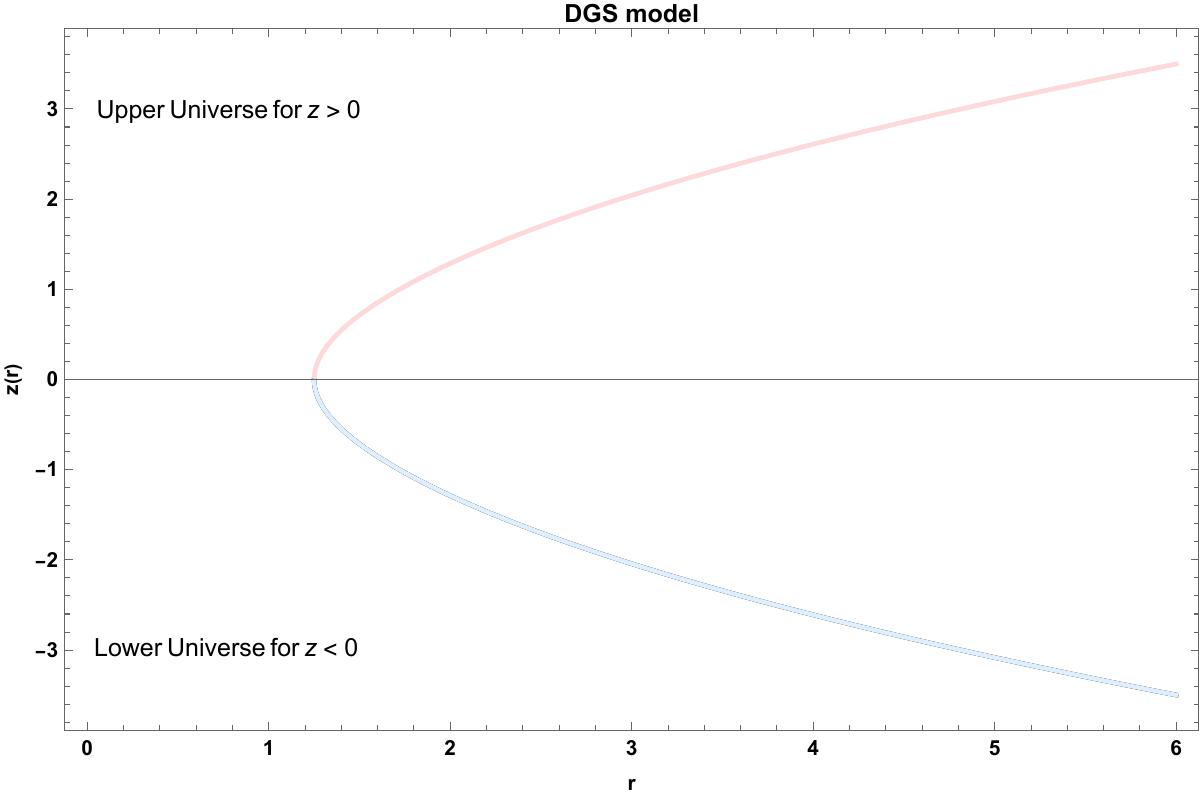}\\
    \includegraphics[width = 8.75 cm, height=6.75cm]{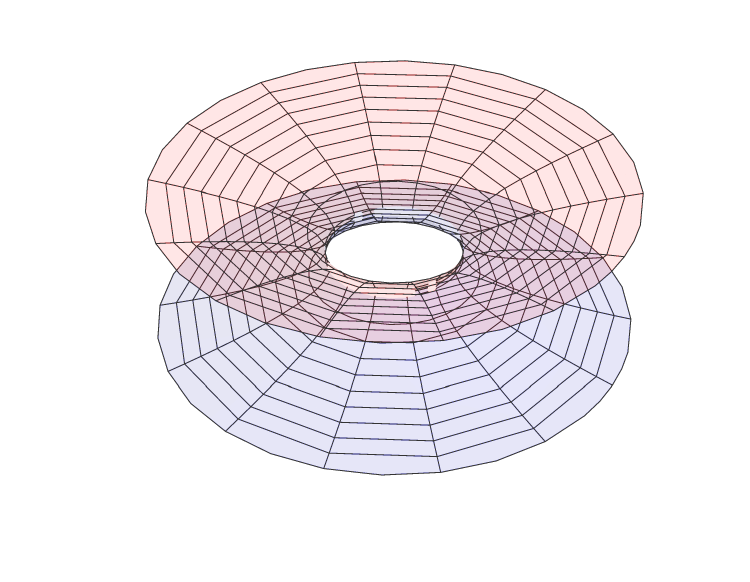}
    \includegraphics[width = 8.75 cm, height=6.75cm]{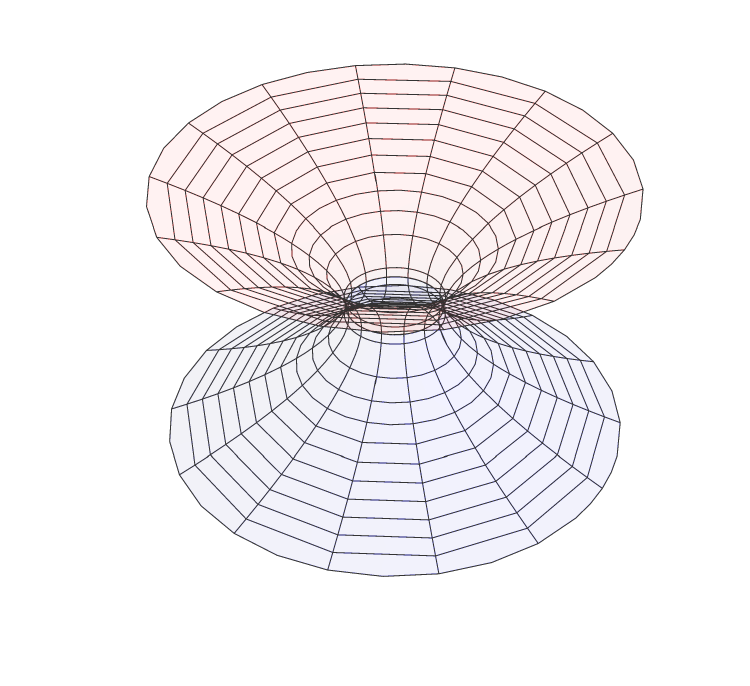}
    \caption{Embedded surface $z(r)$ representing the solution of Eq. \eqref{718}  (top panel) and embedding diagram (bottom panel) for the GUP-corrected Casimir WH for both the KMM and DGS models. We have chosen the following values:  $\lambda = 0.34$, $\Sigma(x) = 1.3$,  $\beta = 0.17$, while $r_0 = 1.7$ (KMM model) or $r_0 =1.25$ (DGS model).}
    \label{fig:2}
\end{figure}

It follows, from Eqs.  \eqref{simplified-2} and \eqref{sh1}, that the radial pressure is given by
 \begin{align}
p_r(r)&=\frac{\pi ^2 }{2160 \lambda }\left[\frac{3 }{r^3}\left(\frac{\lambda -2}{r}+\frac{1}{r_0}\right)+\frac{\beta  \xi_i  }{r^3}\left(\frac{3 \lambda -4}{r^3}+\frac{1}{r_0^3}\right)\right]-\frac{r_0 \Sigma^2(x)}{8 \pi  r^3},
    \label{eq35}
\end{align}
which, owing to Eq. \eqref{pressure-relation}, leads to the tangential pressure expression
\begin{align}
p_t(r)&= \frac{\pi ^2 }{4320 \lambda }\left[\frac{3 }{r^3}\left(\frac{2 \lambda }{r}-\frac{1}{r_0}\right)+\frac{2 \beta  \xi_i  }{r^3}\left(\frac{3 \lambda -1}{r^3}-\frac{1}{2 r_0^3}\right)\right]+\frac{r_0 \Sigma^2(x)}{16 \pi  r^3}.
    \label{eq36}
\end{align}
Equations \eqref{sh1}--\eqref{eq36},  jointly with Casimir energy density  formula \eqref{eq27} and the condition $\Phi(r)=const.$,  completely characterize the WH geometry.  

The embedding diagrams (see Fig.  \ref{fig:2}) can  be traced following the standard procedure outlined in Ref. \cite{Morris1988cz}. First of all, we take advantage of the  symmetries and  consider the equatorial time-fixed two-dimensional slice (cf. Eq. \eqref{metric_RR})
\begin{equation}\label{715}
ds^2=\frac{1}{\Sigma^{2}(x) \left(1-\frac{b(r)}{r}\right)} dr^2+  \frac{r^2}{\Sigma^{2}(x)} d\phi^2.
\end{equation}
After that, we express the line element of the three-dimensional Euclidean embedding space  via cylindrical coordinates ($r, z, \phi$) as 
\begin{equation}\label{716}
ds^2= dr^2 + dz^2 +r^2 d\phi^2.
\end{equation}
Then,   the axially symmetric  embedded surface admits  the metric 
\begin{equation}\label{717}
ds^2= \left[1+ \bigg(\frac{dz}{dr} \bigg)^2 \right]dr^2 + r^2 d\phi^2,
\end{equation}
where $z(r)$ satisfies the following equation, obtained upon comparing Eqs. ($\ref{715}$) and ($\ref{717}$): 
\begin{equation}\label{718}
\frac{dz}{dr}=\pm \left[ \frac{1}{\Sigma^{2}(x)}\left(\frac{1}{ 1-\frac{b(r)}{r}} -1  \right)\right]^{\frac{1}{2}}.
\end{equation}
With these considerations in mind, let us discuss  the physically consistent WH solutions in the next section.

\subsection{Traversability and stability} \label{Sec:stability}

In this section, we explore the interplay between quantum-gravity effects and WH geometry by investigating  how the  GUP parameter $\beta$ influences the traversability and stability of the WH. We assess the first aspect in  Sec. \ref{Sec:flaring-out}, and then deal with the  second by resorting to two different criteria in  Secs. \ref{Sec:first-stability-approach} and \ref{Sec:second-stability-approach}. We conclude the section with a brief analysis of  Casimir WHs comprising no GUP corrections,  which  admit  $\beta=0$ (see Sec. \ref{Sec:beta-zero-case}).

\subsubsection{Flaring-outward condition}\label{Sec:flaring-out}

\begin{figure}[hbt!]
    \centering
    \includegraphics[width = 8.75 cm, height=6.75cm]{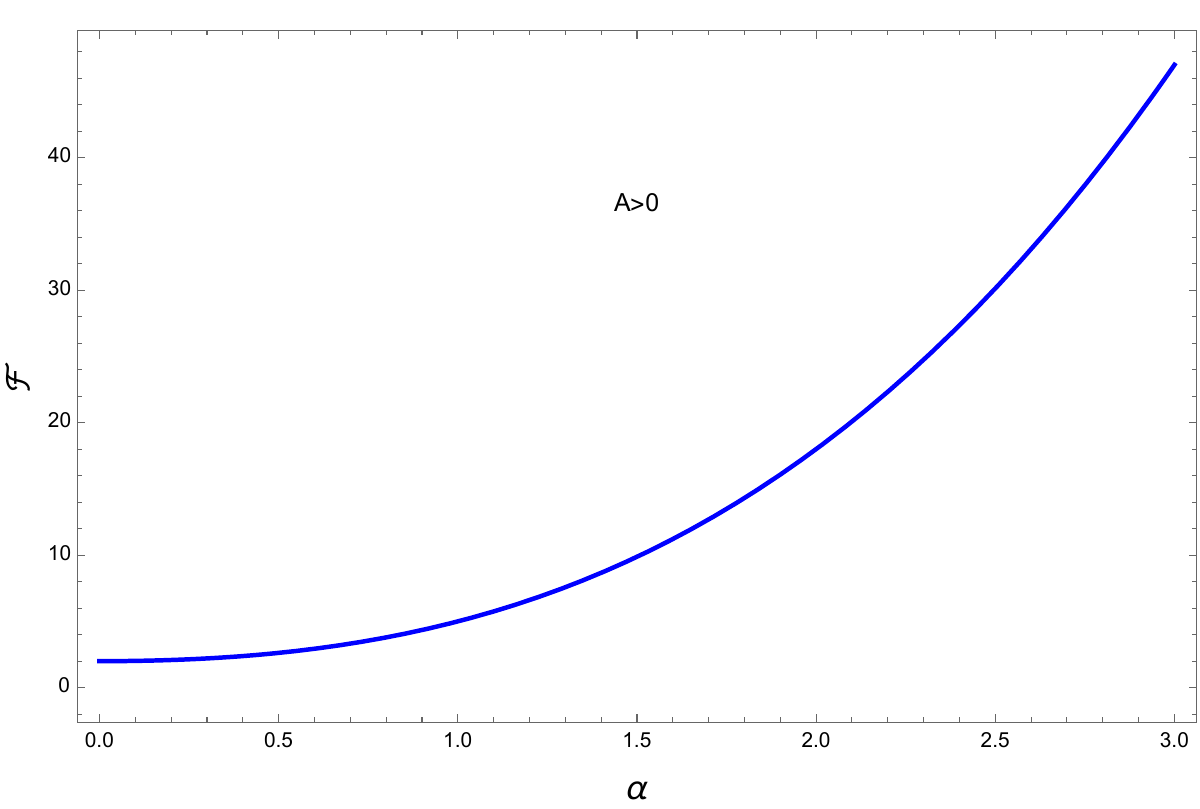}
    \includegraphics[width = 8.75 cm, height=6.75cm]{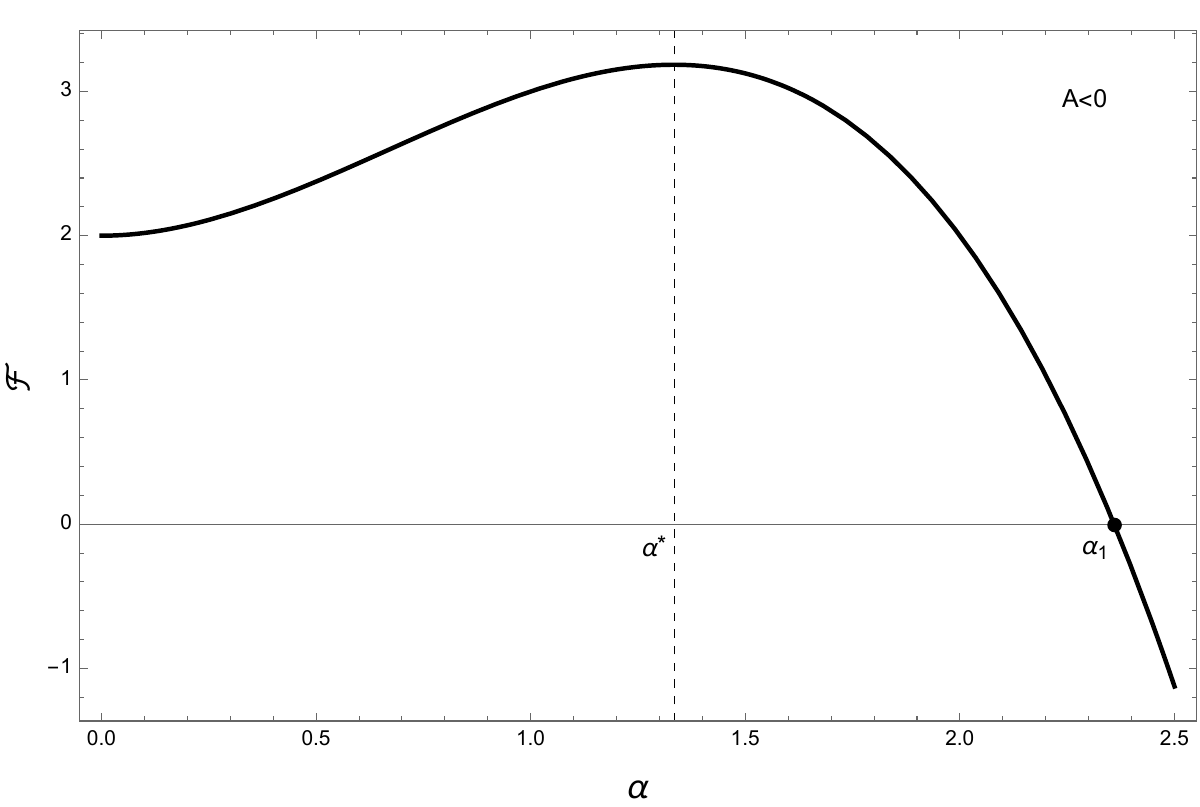}
\caption{Pictorial representation of the $\mathcal{F}(\alpha)$ function defined in Eq. \eqref{F-function} for $A>0$  (left panel) and $A<0$ (right panel). In the first case, the flaring-outward condition \eqref{flaring-out-cond} is always satisfied, while in the second it holds  when $\alpha<\alpha_1$. In this latter scenario, the point $\alpha^*$, where $\mathcal{F}(\alpha)$ attains its maximum, is positive.}
    \label{fig:F-function}
\end{figure}

The WH traversability is related to the fulfillment, near the throat, of the well-known flaring-outward condition \cite{Morris1988cz}
\begin{align}
 b(r)-rb'(r) >0,
 \label{flaring-out-cond}
\end{align}
which  boils down to 
\begin{align}
b'(r_0)<1,
\label{flaring-out-cond-r0}
\end{align}
at $r=r_0$. Starting from Eq.  \eqref{sh1}, we find
\begin{align}
b(r)-rb'(r) = r_0+ \frac{\pi^3 }{270 \lambda  \Sigma^2}\left[3 \left(\frac{2}{r}-\frac{1}{r_0}\right)+\beta  \xi_i  \left(\frac{4}{r^3}-\frac{1}{r_0^3}\right)\right],
\label{flaring-out-cond-expression}
\end{align}
which readily gives at the throat
\begin{align}
b'(r_0)=-\frac{\pi^3 \left(1+\beta \xi_i/r_0^2\right)}{90 \lambda \Sigma^2 r_0^2}.
\label{flaring-out-cond-r0-expression}
\end{align}
The last relation agrees with the upper bound \eqref{flaring-out-cond-r0} provided 
\begin{align}
\lambda>0,    
\label{lambda-positive}
\end{align}
a hypothesis that we henceforth adopt. Now, bearing in mind the  formula \eqref{flaring-out-cond-expression} and upon rescaling the radial variable  according to
\begin{align}
r=\alpha r_0, \qquad \alpha \geq 0,
\label{alpha-variable}
\end{align}
the flaring-outward condition \eqref{flaring-out-cond} is equivalent to requiring that
\begin{align} 
\mathcal{F}(\alpha)>0,
\label{F-function-positive}
\end{align}
where  the real-valued function $\mathcal{F}(\alpha)$ is defined as
\begin{align}
\mathcal{F}(\alpha):= A \alpha^3 + B \alpha^2 + C,    
\label{F-function}
\end{align}
the  coefficients $A$, $B$, and $C$ being given by
\begin{subequations}
\label{coeff-A-B-C}
\begin{align}
A&= 270 \lambda \Sigma^{2} r_0^2 -3 \pi^3 - \pi^3 \beta \xi_i/r_0^2,
\label{A-coefficient}
\\
B&= 6 \pi^3,
\\
C&= 4 \pi^3 \beta \xi_i /r_0^2.
\label{C-coefficient}
\end{align}    
\end{subequations}
A qualitative analysis of the $\mathcal{F} $ function can be given as follows. First of all, from the Descartes rule of signs we find that the algebraic cubic equation 
\begin{align}
\mathcal{F}(\alpha)=0,
\label{cubic}
\end{align}
has  one real root, say $\alpha_1$, and two complex conjugate solutions. In particular, if $A>0$ then $\alpha_1$ is negative, while when $A<0$ $\alpha_1$ is positive (the same result can be obtained using Vi\`ete formulas for cubic equations) \cite{Abramowitz1964}. Additionally, the study of the first-order derivative shows that  $\mathscr{F}(\alpha)$ has a maximum  at $\alpha^*=2B/(-3A)$ and a minimum at $\alpha=0$. The trend of $\mathcal{F}(\alpha)$ is sketched in Fig. \ref{fig:F-function}.

From our preliminary examination, it is possible to conclude  that Eq. \eqref{F-function-positive} is respected for any (positive) $\alpha$ if $A>0$. This  inequality translates into the following upper bound for $\beta$ (cf. Eq. \eqref{A-coefficient}):
\begin{align}
 0<   \frac{\beta \xi_i}{r_0^2} < \frac{3}{\pi^3} \left(90 \lambda \Sigma^{2} r_0^2-\pi^3\right), 
 \label{condition-flaring-1}
\end{align}
which makes sense only if the quantity on the right is  positive, i.e., 
\begin{align}
\Sigma^{2} r_0^2 > \frac{\pi^3}{90 \lambda},
 \label{condition-flaring-2}
\end{align}
where  we recall that we are supposing $\lambda >0$  (cf. Eq. \eqref{lambda-positive}). In other words,  the WH turns out to be traversable when Eqs. \eqref{condition-flaring-1} and \eqref{condition-flaring-2} are satisfied, as these guarantee that Eq. \eqref{flaring-out-cond}, or equivalently Eq. \eqref{F-function-positive},  is fulfilled for any radial distance $r$. 

Let us now consider the scenario where $A$ is negative. In this case,  the WH traversability is ensured  if
\begin{align}
0 \leq \alpha < \alpha_1,
\label{alpha-smaller-alpha1}
\end{align}
and Eq. \eqref{condition-flaring-1} no longer holds, as it  is replaced by
\begin{align}
 \frac{\beta \xi_i}{r_0^2} > \frac{3}{\pi^3} \left(90 \lambda \Sigma^{2} r_0^2-\pi^3\right),
 \label{condition-flaring-3}
\end{align}
while Eq. \eqref{condition-flaring-2} is  still valid. 

At this point, we need to compute the real-valued  root $\alpha_1$ of the cubic \eqref{cubic} \cite{Abramowitz1964}.  By introducing the change of variable 
\begin{align}
\alpha=y-\frac{B}{3A},
\label{alpha-y-variables}
\end{align}
we obtain  the depressed cubic
\begin{align}
y^3+\mathcal{P}y+\mathcal{Q}=0,
\label{depressed-cubic}
\end{align}
having coefficients 
\begin{subequations}
 \begin{align}
\mathcal{P}&=-\frac{B^2}{3A^2},
\\
\mathcal{Q}&=\frac{2B^3+27A^2C}{27A^3},
\label{q-factor}
\end{align}   
\end{subequations}
and discriminant 
\begin{align}
\Delta=-\left(4\mathcal{P}^3+27\mathcal{Q}^2\right)=-\frac{C \left(27 A^2 C +4 B^3\right)}{A^4},
\end{align}
which  always attains  negative values in view of  Eq. \eqref{coeff-A-B-C}. According to the general theory of algebraic third-order equations, this means that the original cubic  \eqref{cubic}  has only one real root, as  pointed out before. Therefore, $\alpha_1$ can be worked out via the Cardano formula jointly with Eq. \eqref{alpha-y-variables}, which gives 
\begin{align}
\alpha_1= \left(u_1\right)^{1/3}+\left(u_2\right)^{1/3}-\frac{B}{3A},
\label{alpha_1 solution}
\end{align}
where
\begin{align}
u_{1,2}=-\frac{\mathcal{Q}}{2} \mp \frac{1}{6}\sqrt{-\frac{\Delta}{3}}.    
\end{align}
It is now evident that $\alpha_1$ is positive when $A<0$, as  both  terms $-\mathcal{Q}/2$ and $-B/(3A)$ are positive in this case.

By performing a detailed investigation, we find that $ \alpha_1$ is always larger than one. In particular, provided that Eq. \eqref{condition-flaring-3} is enforced, we have
\begin{align}
\alpha_1 > \frac{3}{2}. 
\label{alpha1-relation-1}
\end{align}
In addition, 
\begin{align}
\alpha_1 >n >2^{2/3} \approx 1.59, 
\label{alpha1-relation-2}
\end{align}
if the GUP parameter $\beta$ is subject to
\begin{align}
0< \frac{3}{\pi^3} \left(90 \lambda \Sigma^{2} r_0^2-\pi^3\right)<  \frac{\beta \xi_i}{ r_0^2}  < \frac{3 n^2 }{\pi ^3 \left(n^3-4\right)  } \left[90 n \lambda   r_0^2 \Sigma ^2+\pi ^3 (2-n)\right], 
\label{beta-flaring-condition-1}
\end{align}
where we have taken into account the lower bound \eqref{condition-flaring-3} (notice that the above relation is well-defined  due to the fact that Eq. \eqref{condition-flaring-2} is valid and $n>2^{2/3}$). Therefore, thanks to Eqs. \eqref{alpha1-relation-1} and \eqref{alpha1-relation-2} we can conclude that the shape function always satisfies the  flaring-outward condition  \eqref{flaring-out-cond}  near the throat, which is thus not spoiled by the presence of $\beta$.  This is a crucial result, which demonstrates that Casimir WHs  incorporating GUP corrections can be formed in the RR gravity scenario.

\subsubsection{First Stability approach }\label{Sec:first-stability-approach}

The WH stability can be probed using a first stability approach that relies  on the  introduction of  the (squared) adiabatic sound velocity \cite{Sadeghi:2013dea,Luongo:2018lgy,Capozziello:2020zbx} 
\begin{equation}\label{Sound}
v_s^2 = \frac{d \langle p \rangle}{dr} \left( \frac{d\rho}{dr}\right)^{-1},
\end{equation}
where  $\langle p \rangle = \frac{1}{3}(p_r + 2p_t)$ is the pressure averaged across the three spatial dimensions. This quantity embodies the speed at which small disturbances or sound waves propagate through a medium in an adiabatic process (i.e., without the exchange of heat) and hence permits to investigate  how small perturbations affect the WH geometry. Plugging Eqs. (\ref{eq27}), (\ref{eq35}), and (\ref{eq36}) into the formula \eqref{Sound}, we find
\begin{align}
v_s^2 = \frac{2}{3 \lambda }-1.
\label{Sound-2}    
\end{align}
The WH  stability follows from the enforcement of  the causality condition 
\begin{align}
0 \leq v_s^2 < 1,   
\end{align}
which ensures that $v_s^2$ is smaller than the speed of light, as required by the principles of special relativity. Owing to Eq. \eqref{Sound-2}, this requirement is satisfied when
\begin{align}
 \frac{1}{3} < \lambda \leq \frac{2}{3},
 \label{lambda-constraint}
\end{align}
which agrees with the flaring-outward constraint  \eqref{lambda-positive}, thus implying that the WH is both stable and traversable.

Interestingly,  in Eq. \eqref{Sound-2},  we have obtained the same result as in Ref. \cite{Tangphati:2023nwz}, which deals with generic WHs in RR gravity. Thus, by adopting this first  paradigm, $\beta$ does not  influence the WH stability. The situation will change if a different stability criterion is used, as we will describe in the next section. This will allow us to show the pivotal role fulfilled by $\beta$ in the evaluation of the WH properties. Hereafter, we will suppose Eq. \eqref{lambda-constraint}.

\subsubsection{Second stability approach}\label{Sec:second-stability-approach}

The second  stability scheme is based on a mathematical method, first introduced in Ref. \cite{Herrera:1992lwz} and recently refined in Ref. \cite{Abreu:2007ew},  known as {\it Herrera cracking technique}. This procedure   is used to identify the possible appearance of  \qm{crackings}, i.e.,   breaks, in the fluid distributions of self-gravitating compact objects caused by  the presence of local  arbitrarily small pressure anisotropies and density fluctuations. 
The formalism is suitable for analyzing also WH behaviour (see e.g. Refs. \cite{Tayyab2023,ZeeshanGul:2024yhw}) and its 
key result is that a stable system must satisfy specific constraints formulated in terms of the speed of sound. Indeed,  let $\Gamma:=p_t-p_r$ represent the local pressure anisotropy, then the relative order of magnitude of pressure and density perturbations give   
\begin{align}
    \frac{\delta \Gamma}{\delta \rho} \sim \frac{\delta\left(p_t-p_r\right)}{\delta \rho} \sim \frac{\delta p_t}{\delta \rho}-\frac{\delta p_r}{\delta \rho} \sim V_t^2 - V_r^2,
\end{align}
where 
\begin{align}
V^2_t&= \frac{d p_t}{d \rho}, 
\label{tangential-2}
\\
V^2_r&= \frac{d p_r}{d \rho}, 
\label{radial-2}
\end{align}
denote  the transverse and radial components of  the (squared)  sound velocity, respectively. The former quantity measures the response of the object to an outside pressure, while the latter  the changes of  pressure produced by an inside push. In our geometry, Eqs. \eqref{tangential-2} and \eqref{radial-2} yield 
\begin{align}
V^2_t&= \frac{1}{8 \lambda  \left(1+\frac{3 \beta  \xi_i }{2 r^2}\right)} \left[\frac{3 r}{r_0}-8 \lambda +4 r \beta  \xi_i   \left(\frac{1}{4 r_0^3}+\frac{1-3 \lambda }{r^3}\right)-\frac{270 \lambda   \Sigma ^2}{\pi ^3} r r_0 \right],
\\
V^2_r&=  -\frac{1}{4 \lambda  \left(1+\frac{3 \beta  \xi_i }{2 r^2}\right)}\left[\frac{3 r}{r_0}+4 (\lambda -2)+2 r\beta  \xi_i   \left(\frac{1}{2 r_0^3}+\frac{3 \lambda -4}{r^3}\right)-\frac{270 \lambda  \Sigma ^2}{\pi ^3} r r_0\right],
\end{align}
which, bearing in mind Eqs. \eqref{alpha-variable} and \eqref{coeff-A-B-C},  can be written equivalently as
\begin{align}
V^2_t&= -\frac{1}{\lambda  \left(\frac{4 }{3} B \alpha ^2  +3 C\right)} \left[ A \alpha ^3+\frac{4}{3}   \lambda B \alpha ^2+C \left(3 \lambda -1\right)\right],
\label{V-tang-2}
\\
V^2_r&= \frac{1}{\lambda  \left(\frac{2 }{3} B \alpha ^2  +\frac{3}{2} C\right)} \left[ A \alpha ^3+\frac{2}{3} \left(2- \lambda \right) B \alpha ^2+ C \left(2-\frac{3}{2} \lambda\right)\right].  
\label{V-rad-2}
\end{align}

To avoid faster-than-light signals and the related causality violations, both $V^2_t$ and $V^2_r$ must fall within the interval $[0,1)$, i.e.,
\begin{subequations}
\label{causality-Herrera}
\begin{align}
0 \leq V^2_t <1, 
\label{causality-Herrera-Vtang}
\\
0 \leq V^2_r <1,
\label{causality-Herrera-Vrad}
\end{align}
\end{subequations}
which implies $\vert V^2_t - V^2_r \vert <1$ \cite{Herrera:1992lwz}. Then, as demonstrated in  Ref. \cite{Abreu:2007ew},  potentially unstable (resp. stable) regions occur within  anisotropic matter configurations when the (squared) tangential speed of sound  is larger (resp. smaller) than the radial  one. Therefore, we can say that 
\begin{equation}
-1 < V^{2}_{t} - V^{2}_{ r}  < 1  \Rightarrow 
\left\{
\begin{array}{c l}
  -1 < V^{2}_{t} - V^{2}_{ r} \leq 0,    &  \qquad \mathrm{WH\; is \; potentially \; stable}, \\
     &    \\      
0 \leq  V^{2}_{t} - V^{2}_{ r}  < 1,    &  \qquad   \mathrm{WH\; is\; potentially \; unstable}.
\end{array}
\right.
\label{VelocityStability1}
\end{equation}

Bearing in mind Eqs. \eqref{causality-Herrera} and \eqref{VelocityStability1}, we find that the WH is stable if the following inequalities hold: 
\begin{subequations}
\label{stability-conditions}
\begin{align}
0 \leq V_r^2 <1,
\label{stability-conditions-final-1}
\\
V_t^2 \geq 0,
\label{stability-conditions-final-2}
\\
-1 < V_t^2 - V_r^2 \leq 0.
\label{stability-conditions-final-3}
\end{align}   
\end{subequations}

Starting from Eqs. \eqref{V-tang-2} and \eqref{V-rad-2}, we  arrive at a first fundamental conclusion: when $A>0$, the WH is never stable, because both $V^2_r$ and $V^2_t$ do not satisfy the causality conditions \eqref{causality-Herrera}. Therefore, we hereafter suppose $A<0$ and the ensuing relations \eqref{condition-flaring-2} and \eqref{condition-flaring-3}.

\begin{figure}[hbt!]
    \centering
\includegraphics[scale=0.70]{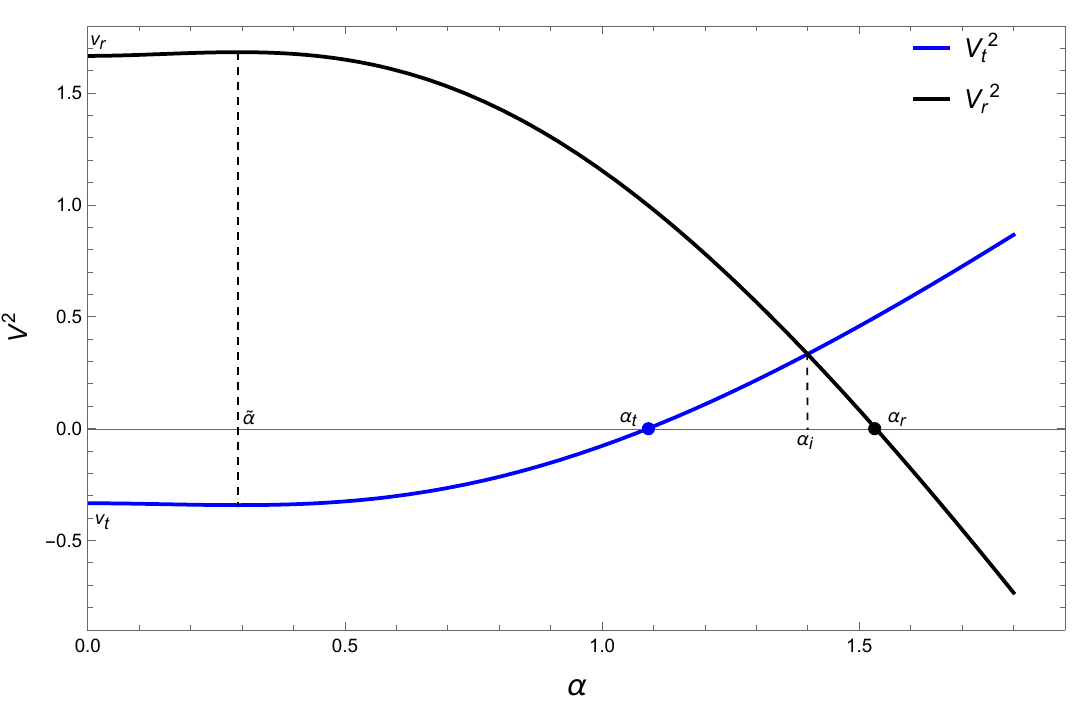}\hspace{1.25cm}
\caption{A pictorial sketch of the functions $V_t^2$ and $V_r^2$, defined in Eqs. \eqref{V-tang-2} and \eqref{V-rad-2}. }
    \label{fig:Vt-Vr-functions}
\end{figure}

To investigate the WH stability, it is essential  to figure out the behaviour of $V_t^2$ and $V_r^2$. A first idea  can be obtained by considering  Fig. \ref{fig:Vt-Vr-functions}, which  provides a pictorial representation of these functions. Let us begin with the study of the derivatives, which   gives
\begin{align}
\frac{d V_t^2}{d \alpha}&= -\frac{12 \alpha  AB }{\lambda  \left(4 \alpha ^2 B+9 C\right)^2} \mathcal{G}(\alpha),
\\
\frac{d V_r^2}{d \alpha}&= \frac{24 \alpha  AB }{\lambda  \left(4 \alpha ^2 B+9 C\right)^2} \mathcal{G}(\alpha),
\end{align}
where we have introduced the real-valued function
\begin{align}
    \mathcal{G}(\alpha):= \alpha ^3+ \frac{27}{4}\frac{  C}{ B} \alpha +2\frac{ C}{A}.
\end{align}
Therefore, the monotonicity of $V_t^2$ and $V_r^2$ is ruled by  the sign of $\mathcal{G}(\alpha)$. First of all, the depressed cubic $\mathcal{G}(\alpha)=0$ admits one real-valued root, which we denote with $\tilde{\alpha}$, and the discriminant
\begin{align}
\tilde{\Delta}= -\frac{27}{16} C^2 \left(\frac{64}{A^2}+\frac{729 C}{B^3}\right).
\end{align}
Since $\tilde{\Delta}$ is  negative, we  can invoke the Cardano formula for evaluating $\tilde{\alpha}$, which thus assumes the form
\begin{align}
\tilde{\alpha}=   \left(-\frac{C}{A}+\frac{1}{6} \sqrt{-\frac{\tilde{\Delta}}{3}}\right)^{1/3}+ \left(-\frac{C}{A}-\frac{1}{6} \sqrt{-\frac{\tilde{\Delta}}{3}}\right)^{1/3}.
\label{tilde-alpha}
\end{align}
In our hypotheses,   $\tilde{\alpha}$ is positive. Then,  from the analysis of the function $\mathcal{G}(\alpha)$, we realize that $\mathcal{G}(\alpha)<0$ if $0 \leq \alpha < \tilde{\alpha}$, and $\mathcal{G}(\alpha)>0$ when $\alpha > \tilde{\alpha}$. This fact permits to conclude that $\alpha=\tilde{\alpha}$ is a minimum for $V_t^2$ and a maximum for $V_r^2$. Moreover, as a consequence of Eq.  \eqref{lambda-constraint} we have
\begin{align}
\nu_t &:= V_t^2 \left(\alpha=0\right)=\frac{1}{3\lambda} \left(1-3\lambda\right) <0, 
\label{nu-t}
\\
\nu_r &:= V_r^2 \left(\alpha=0\right)=\frac{2}{3\lambda} \left(2-\frac{3}{2}\lambda\right) \geq 1.   
\label{nu-r}
\end{align}

At this stage, let us calculate the explicit expressions for the points $\alpha_t$ and $\alpha_r$ where the tangential and radial sound velocities, respectively, vanish. These quantities can be found upon solving the cubics in square brackets in Eqs. \eqref{V-tang-2} and \eqref{V-rad-2}. It follows from the  the Descartes rule of signs that both $\alpha_t$ and $\alpha_r$ are positive, and, after a detailed analysis, we obtain
\begin{align}
\alpha_t &= -\frac{4\lambda}{9}\frac{B}{A} + \left(-\frac{\mathcal{Q}_t}{2} - \frac{1}{6}\sqrt{-\frac{\Delta_t}{3}}\right)^{1/3}+\left(-\frac{\mathcal{Q}_t}{2} + \frac{1}{6}\sqrt{-\frac{\Delta_t}{3}}\right)^{1/3},
\label{alpha-t}
\\
\alpha_r &=\frac{2}{9}\frac{B}{A} \left(\lambda-2\right)+\left(-\frac{\mathcal{Q}_r}{2} - \frac{1}{6}\sqrt{-\frac{\Delta_r}{3}}\right)^{1/3}+\left(-\frac{\mathcal{Q}_r}{2} + \frac{1}{6}\sqrt{-\frac{\Delta_r}{3}}\right)^{1/3},
\label{alpha-r}
\end{align}
with
\begin{align}
\mathcal{Q}_t&= \frac{C(3  \lambda-1) }{ A}+\frac{128 B^3 \lambda ^3}{729 A^3},
 \nonumber \\
\Delta_t &= \frac{C (1-3 \lambda ) }{27 A^4} \left[729 A^2 C (3 \lambda -1)+256 B^3 \lambda ^3\right],
 \nonumber \\
\mathcal{Q}_r&= \frac{C (4-3 \lambda )}{2 A}-\frac{16 B^3 (\lambda -2)^3}{729 A^3} <\mathcal{Q}_t,
 \nonumber \\
\Delta_r &=\frac{C (4-3 \lambda )}{108 A^4}  \left[729 A^2 C (3 \lambda -4)+64 B^3 (\lambda -2)^3\right] <\Delta_t.
\end{align}

Let us now determine the intersection point $\alpha_i$ of the functions $V_t^2$ and $V_r^2$. Starting from Eqs. \eqref{V-tang-2} and \eqref{V-rad-2}, we get
\begin{align}
V_t^2 - V_r^2 = -\frac{9}{\lambda\left(4  B \alpha ^2   +9  C\right) } \mathcal{G}_v(\alpha),
\label{differce-Vtang-Vrad}
\end{align}
where
\begin{align}
    \mathcal{G}_v(\alpha):=  A \alpha ^3 +\frac{8}{9} B \alpha ^2 + C.
\end{align}
Therefore,  $\alpha_i$ represents the positive real-valued solution of the cubic $\mathcal{G}_v(\alpha)=0$, which yields 
\begin{align}
\alpha_i &= -\frac{8 B}{27 A} + \left(-\frac{\mathcal{Q}_i}{2} - \frac{1}{6}\sqrt{-\frac{\Delta_i}{3}}\right)^{1/3}+\left(-\frac{\mathcal{Q}_i}{2} + \frac{1}{6}\sqrt{-\frac{\Delta_i}{3}}\right)^{1/3},
\label{alpha-i}
\end{align}
with
\begin{align}
\mathcal{Q}_i&= \frac{1}{27 A^3}\left(27 A^2C+\frac{1024 B^3}{729}\right),
 \nonumber \\
\Delta_i &= -\left(\frac{2048 B^3 C}{729 A^4}+\frac{27 C^2}{A^2}\right).
\end{align}

We are now ready to address the question of the WH stability. First of all,  the physically meaningful solutions of the system \eqref{stability-conditions} are those for which  $\alpha_r > \alpha_t$, i.e., $\alpha_r$ must be strictly larger than $\alpha_t$. For if this  relation were not true, then $V_r^2$ and $V_t^2$ would not satisfy causality criteria \eqref{causality-Herrera} at the same time. In our hypotheses, the inequality $\alpha_r > \alpha_t$ is satisfied  if
\begin{align}
-\frac{\mathcal{Q}_r}{2}  -\frac{1}{6}\sqrt{-\frac{\Delta_r}{3}}>-\frac{\mathcal{Q}_t}{2}-\frac{1}{6}\sqrt{-\frac{\Delta_t}{3}},
\label{zeros-constraint-1}
\end{align}
which gives
\begin{align}
  \Delta_r - \Delta_t +3 \left(\mathcal{Q}_r-\mathcal{Q}_t\right) \left(9 \mathcal{Q}_r-9 \mathcal{Q}_t-2 \sqrt{3} \sqrt{-\Delta_t}\right) >0.
  \label{zeros-constraint-2}
\end{align}
A necessary but not sufficient condition for the above relation to hold is that $9 \mathcal{Q}_r-9 \mathcal{Q}_t-2 \sqrt{3} \sqrt{-\Delta_t}<0$, a constraint that is always fulfilled in our hypotheses\footnote{We observe that Eq. \eqref{zeros-constraint-1} is automatically true if Eqs. \eqref{condition-flaring-1} and \eqref{condition-flaring-2},  i.e. the conditions for $A>0$, are valid. However, we know that when  $A>0$ the WH cannot be stable. }. Given these premises, a  key result   is that (cf. Eq. \eqref{nu-r})
\begin{align}
    V^2_r\left(\alpha=\tilde{\alpha}\right) \geq \nu_r \geq 1,
\end{align}
which translates into the fact that $V^2_r$ complies with the causality condition \eqref{stability-conditions-final-1}  only in the interval 
\begin{align}
\tilde{\alpha} < \alpha \leq \alpha_r.
\label{causality-Vr}
\end{align}
Moreover, the behaviour of the function $\mathcal{G}_v(\alpha)$ occurring in Eq. \eqref{differce-Vtang-Vrad} permits to figure out another crucial point. Indeed, $\mathcal{G}_v(\alpha)$ is positive when $0 \leq \alpha < \alpha_i$ and attains a maximum at $\alpha=-16B/(27A)$. Therefore, we can claim that  $V_t^2 - V_r^2  \leq 0$ when 
\begin{align}
0 \leq \alpha \leq \alpha_i.
\label{Vt-Vr-comparison}
\end{align}
Therefore, bearing in mind  Eqs. \eqref{causality-Vr} and \eqref{Vt-Vr-comparison}, we can conclude that the system \eqref{stability-conditions} is satisfied if (see also Fig. \ref{fig:Vt-Vr-functions})
\begin{subequations}
\label{system-1}
\begin{align}
\tilde{\alpha}<\alpha \leq \alpha_r,
\label{system-1-a}
\\
\alpha \geq \alpha_t,
\label{system-1-b}
\\
0 \leq \alpha \leq \alpha_i.
\label{system-1-c}
\end{align}
\end{subequations}
Now, since $\alpha_t < \alpha_i <\alpha_r$ and assuming that  $\alpha_r > \alpha_t$ (which implies that Eq. \eqref{zeros-constraint-1},  or equivalently Eq. \eqref{zeros-constraint-2},  is valid),  the system \eqref{system-1} simplifies to
\begin{align}
\alpha_t \leq \alpha \leq \alpha_i.
\label{system-2}
\end{align}
Therefore, we can claim that the WH is stable (and traversable, see below) if Eq. \eqref{system-2} is met. Moreover, $\alpha_i$ is subject to the key  relation
\begin{align}
    \alpha_i >\frac{3}{2},
\end{align}
which proves that in the vicinity of the throat the WH is stable. Notice that also $\alpha_1$ satisfies an analogous inequality (see Eq. \eqref{alpha1-relation-1})\footnote{It is interesting to note that in  general, given any integer $m>1$, the condition $\alpha_i > m $ is satisfied according to  the pattern
\begin{equation*}
\left\{
\begin{array}{c l}
  9 A m^3+8 B m^2+9 C > 0,    &  \qquad {\rm if}\;  m\;{\rm mod}  \,3\neq 0, \\
     &    \\      
A m^3+8 B \left(\frac{m}{3}\right)^2+C >0,    &  \qquad   {\rm if}\;  m\;{\rm mod}  \,3= 0.
\end{array}
\right.
\end{equation*} }.

In order to investigate the WH traversability and stability,  Eq. \eqref{system-2} should be compared with the constraint \eqref{alpha-smaller-alpha1}. Bearing in mind formulas \eqref{alpha_1 solution} and \eqref{alpha-i}, we arrive at the fundamental inequality
\begin{align}
\alpha_1 > \alpha_i,
\label{crucial-result}
\end{align}
which leads to the crucial conclusion that the GUP-corrected Casimir WH can be both traversable and stable. Indeed, the validity of  Eq. \eqref{crucial-result} implies that when the WH is stable (i.e., Eq. \eqref{system-2} is valid)   then  it is also traversable (i.e., Eq. \eqref{alpha-smaller-alpha1} holds), but the converse is not true. This result is also physically reasonable, as the requirement of stability reduces the allowed range for $\alpha$ from the original interval \eqref{alpha-smaller-alpha1},  where the flaring-outward condition \eqref{flaring-out-cond} is respected, to the stricter one \eqref{system-2}. This reflects the fact that stability is an additional  requirement that further constrains the WH geometry. 

It is worth noticing that  $\alpha_1<\alpha_r$ when $1/3 < \lambda <1/2$, while $\alpha_1>\alpha_r$ for $1/2 \leq \lambda \leq 2/3$. Then, owing to Eq. \eqref{crucial-result}, the situation is  the following:
\begin{align}
 0<   \tilde{\alpha}<\alpha_t <\alpha_i <\alpha_1 <\alpha_r, \qquad {\rm if} \; \; \frac{1}{3} < \lambda < \frac{1}{2},
\end{align}
and
\begin{align}
 0<   \tilde{\alpha}<\alpha_t <\alpha_i <\alpha_r <\alpha_1, \qquad {\rm if} \; \; \frac{1}{2} \leq \lambda \leq  \frac{2}{3},
\end{align}
where we have assumed  Eq. \eqref{zeros-constraint-1}, or equivalently Eq. \eqref{zeros-constraint-2}, to suppose that  $\alpha_t <\alpha_r$.

To further stress how the GUP parameter $\beta$ influences the WH stability and traversability, in the next section we will investigate Casimir WHs in the regime $\beta=0$. Interesting differences with the model analyzed so far will emerge.

\subsubsection{The case $\beta=0$}\label{Sec:beta-zero-case}

The model with  $\beta=0$ refers to Casimir WHs involving no GUP corrections. In this context,  starting from Eq. \eqref{sh1}, we obtain
\begin{align}
(b-rb')_{\beta=0}=    r_0-\frac{\pi ^3 (\alpha -2)}{90 \lambda \alpha    r_0 \Sigma ^2},
\end{align}
and hence the flaring-outward condition \eqref{flaring-out-cond} is secured, in the hypothesis $\lambda>0$, if either the constraint \eqref{condition-flaring-2} is valid or the following relations are met:
\begin{equation}
\left\{
\begin{array}{c l}
 90 \lambda r_0^2 \Sigma^{2} < \pi^3, \\
     &    \\      
\alpha < \frac{2\pi^3}{\pi^3-90\lambda r_0^2 \Sigma^{2}}.  
\end{array}
\right.
\label{flaring-beta=0}
\end{equation}

The  (squared) adiabatic sound velocity retains the same value as in Eq. \eqref{Sound}, which means that the inequality \eqref{lambda-constraint} guarantees WH stability according to the first stability approach outlined in Sec. \ref{Sec:first-stability-approach}.  Following the recipe of the second stability method described in Sec. \ref{Sec:second-stability-approach} and upon assuming Eq. \eqref{lambda-constraint}, we can now  calculate the transverse and radial sound velocities. Starting from Eqs.   \eqref{tangential-2} and \eqref{radial-2}, we arrive at the following formulas:
\begin{align}
\left(V_t^2\right)_{\beta=0} &=  \frac{3}{8} \alpha  \left(\frac{1}{\lambda }-\frac{90 r_0^2 \Sigma ^2}{\pi ^3}\right)-1,
\\ \nonumber
\left(V_r^2 \right)_{\beta=0} &= \frac{1}{\lambda }\left(2-\frac{3 \alpha }{4}\right)+\frac{135 \alpha  r_0^2 \Sigma ^2}{2 \pi ^3}-1,
\end{align}
which permit to achieve a first  interesting conclusion: if Eq. \eqref{condition-flaring-2} holds, then   the WH is traversable but not stable,  as the stability requirements \eqref{stability-conditions} are not satisfied. This is in sharp contrast  to the scenario with $\beta \neq 0$, where  Eq. \eqref{condition-flaring-2} does not compromise either  WH stability or traversability. On the other hand, in the setup with $\beta=0$, the WH is both stable and traversable  if $\lambda<2/3$ and 
\begin{align}
    \frac{8 \pi ^3 \lambda }{3 \left(\pi ^3-90 \lambda  r_0^2 \Sigma ^2\right)}\leq \alpha \leq \frac{16 \pi ^3}{9 \left(\pi ^3-90 \lambda  r_0^2 \Sigma ^2\right)},
\end{align}
which is consistent with Eq. \eqref{flaring-beta=0}.  This   situation  resembles the previous framework  (see Eq. \eqref{system-2}), albeit with different assumptions (recall that Eq. \eqref{condition-flaring-2} does not imply stability when  $\beta=0$). Our analysis thus reveals the remarkable point that the factor $\beta$ can completely alter the criteria for  WH stability. In other words, the fact that the presence of a nonvanishing GUP parameter does not spoil either the traversability or the stability of Casimir WHs is not trivial, as it is not something to be expected automatically from the beginning.

\subsection{The emergence of exotic matter } \label{Sec:exotic-matter}

As pointed out before, the occurrence of exotic matter in WH spacetimes implies the breach of the NEC. We now determine whether the GUP-corrected Casimir WHs in the RR gravity theory adhere to the NEC, which, we recall,  translates into the relations  $\rho(r)+p_r(r) \geq 0$ and  $\rho(r)+p_t(r) \geq 0$ \cite{visser1995lorentzian}.

Bearing in mind Eq. (\ref{eq14}) (or equivalently Eq. \eqref{eq27}) and Eqs. \eqref{eq35} and \eqref{eq36}, we obtain
\begin{align}
\rho+p_r&=-\frac{\mathcal{F}(\alpha)}{2160 \pi   \lambda  r_0^4 \alpha ^6},\label{eq37a}\\
\rho+ p_t &=\frac{1}{4320 \pi   \lambda  r_0^4 \alpha ^6}\left(A \alpha^3 -\frac{C}{2}\right),  \label{eq37b}
\end{align}
where the $\mathcal{F}(\alpha)$ function and its coefficients have been defined in Eqs. \eqref{F-function} and \eqref{coeff-A-B-C}, respectively. It follows from Eq. \eqref{eq37a} that  as soon as the flaring-outward condition \eqref{flaring-out-cond} is met,  $\rho+p_r$ is negative owing to Eq. \eqref{F-function-positive}. Furthermore, we recall that, according to the analysis of  Sec. \ref{Sec:second-stability-approach}, the WH is stable only in the case $A<0$, and, in this hypothesis,  $\rho + p_t$ is also negative. Since we have proved that WH stability implies  traversability (see comments below Eq. \eqref{crucial-result}), we can conclude that traversable and stable GUP-corrected Casimir WHs always require the existence of NEC-violating fields.

The total amount $\mathcal{M}$ of exotic matter residing in the spacetime can be calculated via the volume integral \cite{Visser:2003yf} (see also e.g. Ref. \cite{Jusufi:2020rpw})
\begin{equation}
\mathcal{M}=\oint \left(\rho(r)+p_r(r)\right) dV= 2 \int \limits_{r_0}^{\infty} \left(\rho(r)+p_r(r)\right)dV = 8 \pi  \int \limits_{r_0}^{\infty} \left(\rho(r)+p_r(r)\right) r^2 dr,
\end{equation}
where we have used  the relation $\oint dV = 2\int\limits_{r_0}^{\infty} dV = 8\pi \int\limits_{r_0}^{\infty} r^{2}dr$.  By introducing a cutoff parameter  $\R$, we thus obtain
\begin{align}
\mathcal{M}&= 8 \pi  \int \limits_{r_0}^{\R} \left[\rho(r)+p_r(r)\right] r^2 dr = 8 \pi  \int \limits_{1}^{\alpha_\R} \left[\rho(\alpha)+p_r(\alpha)\right] r_0^3 \alpha^2 d\alpha =
 \nonumber \\
&=-\frac{1}{810 \alpha_\R^3 \lambda  r_0} \left[ \alpha_\R^3 (3 A \log \alpha_\R +3 B+C)-3B \alpha_\R^2 - C \right],
\label{exotic-mass-1}
\end{align}
with $ \alpha_\R = \R / r_0 $. Notably, the GUP factor significantly influences the  exotic mass content. To demonstrate this point, we first evaluate  the behaviour of $\mathcal{M}$ near the throat by expanding  the formula \eqref{exotic-mass-1} around $\alpha_\R=1$:
\begin{align}
    \mathcal{M}= - \frac{ \pi ^3+ 90 \lambda   \Sigma ^2 r_0^2+\pi ^3 \beta  \xi_i/r_0^2}{90 \lambda  r_0} \left(\alpha_\R-1\right)+O\left[\left(\alpha_\R-1\right)^2\right],
\end{align}
and then compare it with the corresponding quantity computed in the regime $\beta=0$:
\begin{align}
      \left(\mathcal{M}\right)_{\beta=0}= -  \frac{ \pi ^3+ 90 \lambda   \Sigma ^2 r_0^2}{90 \lambda  r_0} \left(\alpha_\R-1\right)+O\left[\left(\alpha_\R-1\right)^2\right].
\end{align}
We thus see that, in a region close to the throat, the following inequality holds:
\begin{align}
    \vert \mathcal{M} \vert > \vert  \left(\mathcal{M}\right)_{\beta=0} \vert ,
\end{align}
which entails that the amount of exotic matter  sourcing   GUP-corrected Casimir WHs increases as the value of $\beta$ grows. This is not a novel feature in this research field, as exotic matter exhibits  a similar trend also in Ref. \cite{Sahoo:2023dus}, which deals with   GUP-modified Casimir WHs in symmetric teleparallel gravity. Such result is  physically reasonable, since the larger is $\beta$ the greater is the modulus of the Casimir energy density $\rho(r)$ (cf. Eq. \eqref{eq27}),  implying a more significant presence of NEC-violating fields.  Thus, our  findings confirm that the GUP-modified Casimir energy can serve as a source of exotic matter, which remains a key element for the formation and maintenance of stable WHs.

\subsection{Light deflection angle} \label{Sec:light-deflection}

In this section, we investigate the phenomenon of light deflection in the WH gravitational field.

It follows from Eq. \eqref{metric_RR} that the orbit of a photon in the equatorial plane of the WH spacetime is characterized by the relation
\begin{align}\label{eq50}
    dt^{2}= \frac{1}{ 1-\frac{b(r)}{r}}dr^2+r^{2}d\phi^{2},
\end{align}
where we have introduced the coordinate rescaling $e^{2\Phi(r)}dt^{2} \rightarrow dt^{2}$ taking advantage of the fact that for a zero-tidal-force WH $e^{2\Phi(r)} = const$.

In order to compute the  deflection angle $ \Theta$ of a light ray passing by the WH, we employ a recent geometric approach that relies on the Gauss-Bonnet theorem  \cite{Arakida:2017hrm}. Thanks to  this method,  $ \Theta$ can be written via the integral expression
\begin{align}\label{eq56}
    \Theta = - \int\limits_{0}^{\pi} \,  \int\limits_{r=\frac{\ell}{\sin\phi}}^{\infty} \mathcal{K}dS,
\end{align}
where $\mathcal{K}$ is the Gaussian optical curvature, $dS$ the optical surface element, and  $\ell$  the impact parameter, which, to first approximation, can  be evaluated as the distance of  closest approach to the WH center.  By rearranging  the optical metric (\ref{eq50}) as 
\begin{align}
    dt^{2}= du^2+\mathcal{H}^{2}(u)d\phi^{2},
\end{align}
where $\mathcal{H}=r$ and the new coordinate $u$ satisfies 
\begin{align}
 du= \frac{1}{\sqrt{ 1-\frac{b(r)}{r}}}dr, 
\end{align}
we can write $\mathcal{K}$ through the general formula \cite{Gibbons:2008rj} (see also Eq. (47) in Ref. \cite{Jusufi:2018waj}) 
\begin{align}
\mathcal{K}= -\frac{1}{\mathcal{H}(u)}\left[\frac{dr}{du}\frac{d}{dr}\left(\frac{dr}{du}\right)\frac{d\mathcal{H}}{dr}+\left(\frac{dr}{du}\right)^2\frac{d^2\mathcal{H}}{dr^2}\right],
\end{align}
which in our geometry yields 
\begin{align}\label{eq60}
\mathcal{K} = -\frac{r_0}{2 r^3} + \frac{\pi ^3 }{540 \lambda   \Sigma^{2} r_0^3 r^6 }\left[3 r^2 r_0^2 (r-2 r_0) + \beta  \xi_i  \left(r^3-4 r_0^3\right)\right].
\end{align}
\begin{figure}[htbp!]
    \centering
    \includegraphics[width = 12.75 cm, height=8.75cm]{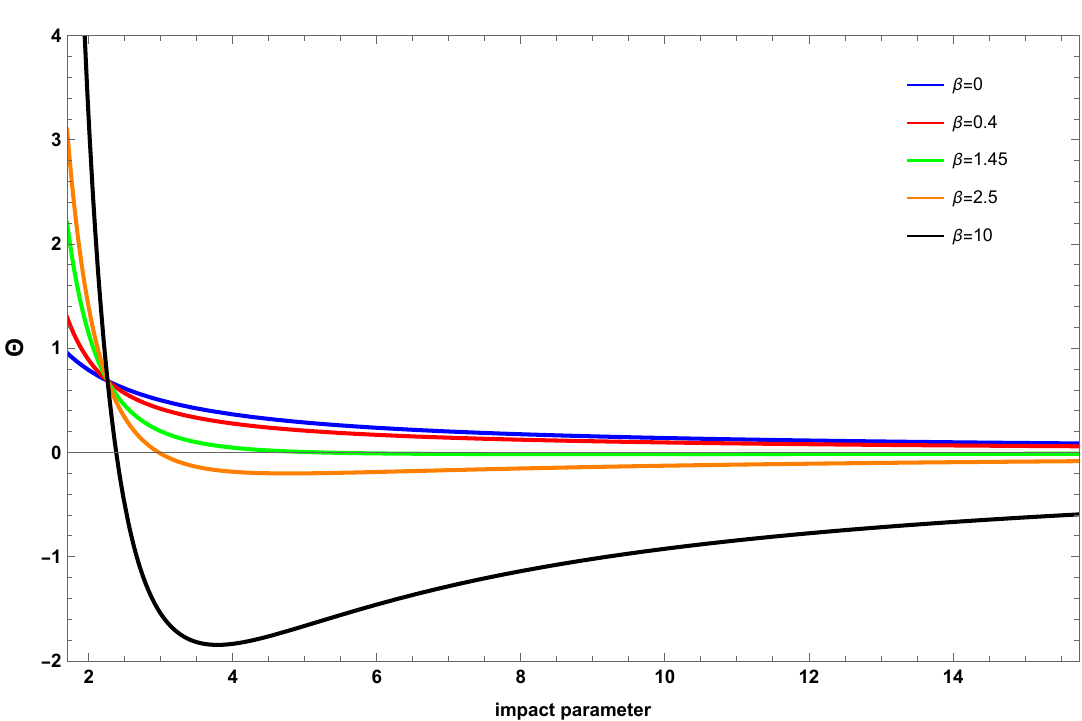}
    \caption{Deflection angle $\Theta$ as a function of the impact parameter $\ell$  for the KMM model with $r_0 = 1.7$, $\lambda = 0.34$,  $\Sigma = 1.7$, and   $\beta$ attaining five different values: $\beta = 0, 0.4, 1.45, 2.5, 10$. The behaviour of $\Theta$ varies depending on whether $\beta$ is zero or non-zero. }
    \label{fig:9}
\end{figure}
We can thus work out the deflection angle by plugging  Eq. (\ref{eq60}) into the expression (\ref{eq56}) and then computing  the ensuing  integral. In this way, we find  the exact result 
\begin{align}
\Theta = \frac{r_0}{\ell}-\frac{\pi ^3}{90 \ell r_0 \lambda   \Sigma^{2}} \left[ 1-\frac{\pi  r_0}{4 \ell}+\beta  \xi_i  \left(\frac{1}{3 r_0^2}-\frac{\pi  r_0}{16 \ell^3}\right)\right].
\label{deflection}
\end{align}

Apart from the contributions associated with the WH geometry,  the deflection angle is influenced by the GUP parameter $\beta$, as indicated by the last term in Eq. \eqref{deflection}. Moreover,  our formula for $\Theta$ reduces to the corresponding GR relation pertaining to zero-tidal-force GUP-corrected Casimir WHs  in the limit $\lambda=\Sigma^2=1$ (see Ref. \cite{Jusufi:2020rpw} for further details).

The deflection angle goes to zero as the impact parameter  approaches  infinity,  indicating that the trajectories of photons moving far away from the WH remain almost unaffected.  Conversely,  the expansion of Eq. \eqref{deflection} around $\ell=r_0$ yields
\begin{align}
\Theta= 1-\frac{\pi ^3 }{90 \lambda  r_0^2 \Sigma ^2} \left[1-\frac{\pi }{4}+\frac{(16-3 \pi ) \beta  \xi_i }{48 r_0^2}\right]  - \left\{\frac{1}{r_0}+\frac{\pi ^3 }{1080 r_0^5 \lambda   \Sigma ^2}\left[ 6   r_0^2(\pi-2)+  \beta  \xi_i (3 \pi -4) \right]\right\}\left(\ell-r_0\right)+O\left[(\ell-r_0)^2\right],
\end{align}
and hence $\Theta$ can be either positive or negative when $\ell$ takes values close to the WH throat.  In the first scenario,   light rays fall toward the WH, while, in the second,    they bend outward  (recall however that to describe the light bending one should consider the absolute value of $\Theta$). Interestingly, there exist situations where the features of $\Theta$ can be somewhat  modified by the presence of a nonvanishing GUP parameter $\beta$. This is evident from  Fig. \ref{fig:9}, which shows that in the regime $\beta=0$ the deflection angle  is always positive, while it   can assume  negative values when $\beta$  is nonvanishing.

\section{Concluding remarks }\label{sec4}

In this paper, we have studied  static and spherically symmetric GUP-modified Casimir WHs within the context of RR gravity, which arises from  incorporating  the energy-dependent metric featuring  Rainbow scenario into the Rastall model (see Sec. \ref{sec2}). These objects are supported by the negative Casimir energy density containing  GUP contributions (see Sec. \ref{Sec.III}) and naturally allow for a quantum-gravity-inspired NEC violation, a possibility that makes them particularly relevant to the current theoretical literature. The GUP has been formulated to modify  the Heisenberg uncertainty principle via the deforming factor $\beta$, which is 
expected to emerge from candidate theories of quantum gravity. GUP corrections alter the overall energy-momentum distribution of matter sourcing  WHs, thereby affecting their geometric properties. This means that the analysis of WHs comprising GUP adjustments offers  precious insights into the role that   quantum gravity can assume  in  the formation of nontrivial spacetime structures. 

In the realm of WHs, three essential questions stand out: traversability, stability, and the existence of exotic matter. We have provided an original and thorough examination of these aspects  in Sec. \ref{Sec:Casimir-WHs}. In particular, we have tackled the stability issue by employing two different paradigms, with the adiabatic sound velocity as the primary physical quantity considered. Notably, we have utilized the so-called Herrera cracking technique, which, despite its elegance and effectiveness, appears to be not widely harnessed in the WH research field. To the best of our knowledge, while the original scheme developed  in Ref.  \cite{Herrera:1992lwz} has been somewhat explored in WH studies  (see e.g. Refs. \cite{Tayyab2023,ZeeshanGul:2024yhw}), its refinement recently worked out in Ref. \cite{Abreu:2007ew}  has not yet been applied to WH stability analysis. Unlike the approach discussed in Sec. \ref{Sec:first-stability-approach}, where the GUP term $\beta$ plays no role in the assessment of the WH stability, the (improved) Herrera cracking procedure has the advantage of highlighting the crucial contribution of  $\beta$.   In fact, by means of this formalism, we have established that there   are always values of  $\beta$ that guarantee  stability, a requirement that in our setup also ensures the fulfillment of the flaring-outward constraint (see Sec. \ref{Sec:second-stability-approach}). In other words, our investigation has allowed us to demonstrate that Casimir WHs framed within the RR pattern can achieve both stability and traversability. Moreover, we have found that these properties are inherently linked to the occurrence of NEC-violating matter, whose  amount increases as  $\beta$ grows.

The proof that Casimir WHs can be  stable is not a result to be taken for granted, as it could potentially be spoiled by the presence of the GUP parameter. In fact, as outlined in Sec.  \ref{Sec:beta-zero-case},  the  stability criteria differ based on whether $\beta$ is  zero or nonzero. This fact   shows that  WHs incorporating GUP corrections are fascinating solutions which deserve further consideration. The main claim  of  this paper is  that such WHs can indeed arise within RR  theory. This represents a key finding especially if we compare it with the outcome of Ref. \cite{Tangphati:2023nwz}, where it is proved that generic zero-tidal-force WHs cannot be formed in RR framework upon considering specific parameter combinations  and equations of state. 

Finally, another fundamental facet of this paper is   the application of the revised Herrera approach and the ensuing stability conditions that we have computed  in Eq. \eqref{stability-conditions}. Given their broad applicability,  we deem that the plethora of WH prototypes devised in the literature should be analyzed exploiting this method to assess whether they are   stable. In this way, it is likely that valuable perspectives in this field will be revealed.

\section*{Acknowledgements}

EB and SC acknowledge the support of INFN {\it sez. di Napoli}, {\it iniziative specifiche QGSKY} and {\it MOONLIGHT2}. AE thanks the National Research Foundation of South Africa for the award of a postdoctoral fellowship.
This paper is based upon work from COST Action CA21136 Addressing observational tensions in cosmology with systematics and fundamental physics (CosmoVerse) supported by COST (European Cooperation in Science and Technology).

\bibliography{references}

\end{document}